\documentclass{svproc}
\usepackage{url}

\usepackage{epsfig}
\usepackage{subfig}
\usepackage{graphicx}
\usepackage{color}
\usepackage[ruled,vlined]{algorithm2e}
\usepackage{epstopdf}
\usepackage{wrapfig}

\begin{document}
\mainmatter              
\title{Computational Thinking in {\em Patch}}
\titlerunning{Computational Thinking in Patch}  
%
\author{Hasan M. Jamil}
\authorrunning{Hasan Jamil} 
%
\tocauthor{Hasan Jamil}
\institute{Department of Computer Science\\ University of Idaho, USA\\
\email{jamil@uidaho.edu}
}

\maketitle              

\begin{abstract}
With the future likely to see even more pervasive computation, computational thinking (problem-solving skills incorporating computing knowledge) is now being recognized as a fundamental skill needed by all students. Computational thinking is  conceptualizing as opposed to programming, promotes natural human thinking style than algorithmic reasoning, complements and combines mathematical and engineering thinking, and it emphasizes ideas, not artifacts. In this paper, we outline a new visual language, called {\em Patch}, using which students are able to express their solutions to eScience computational problems in abstract visual tools. Patch is closer to high level procedural languages such as C++ or Java than Scratch or Snap! but similar to them in ease of use and combines simplicity and expressive power in one single platform.
\end{abstract}

\section{Introduction}

The differences between computing and computational thinking are significant and can be explained in a number of ways. According to Jeannette Wing \cite{Wing06}, ``computational thinking confronts the riddle of machine intelligence: What can humans do better than computers, and what can computers do better than humans? Most fundamentally, it addresses the question: What is computable? Today, we know only parts of the answers to such questions." In particular, computational thinking is 1) conceptualizing, not programming, 2) a way that humans, not computers, think, 3) complements and combines mathematical and engineering thinking, and 4) it is ideas, not artifacts."

In many countries, including USA, teachers, practitioners, developers and researchers are clamoring for new tools and systems to help realize these goals as part of emerging K-12 education standards such as NGSS \cite{NGSS} to better prepare students for new age of STEM education. Computing systems such as Alice, Scratch and Snap!, and educational systems such as Khan Academy (https://www.khanacademy.org/), ALEKS (https://www.aleks.com/) and so on have been developed in response to these pressing needs. Not only these systems and softwares are gaining popularity \cite{ChangLC13s,Saez-LopezRV16}, they are also being accepted by formal academic institutions \cite{ReddyH13} as standards they are able to rely on. They, however, are not above criticisms in the press (e.g., https://tinyurl.com/m8vko7v and https://tinyurl.com/k4lrqmn) and complaints (https://tinyurl.com/kzm5mny) about their delivery and conducts exists, as well as there exist significant concerns among the academicians about their effectiveness \cite{Simonova14s,Farag12s}.

NGSS calls for better integration and use of computing in STEM experiences, and in developing an effective computational thinking repertoire. Differently from computing, computational thinking demands the ability to string reasoning, analytics, knowledge, information, models, and hypotheses in a coherent manner, and often when one of the components is missing or inadequate, to develop one to solve a STEM problem at hand. If the missing component can or need to be programmed, it is prudent to expect a computational thinking approach to computing solutions as well to avoid any impedance mismatch across components within a larger system supporting NGSS or computational thinking. Approach to computational thinking for computing, however, is a lower level concept and appears to have a discipline specific bias as to how abstract should it be.

So, it is not a surprise that the computational physicists (e.g., \cite{LandauMHBKB14}) insist on a more hands on computer programming experience than computational biologists (e.g., \cite{Elhai11}) advocating visual or NLP based programming, or computational thinking.  Fortunately, millions of K-12 students are now conversant in the graphical language Scratch \cite{CalaoMCR15s} (developed at MIT). However, the language is incredibly clunky if one wants to go beyond what it was designed to do (make simple games), but a derivative language, Snap! \cite{LeleuxHSYSG15s} (developed at UC Berkeley), expands its horizons, making functions practical, enabling object oriented programming, and fixing other deficiencies of Scratch. Yet, many educators believe that it is a fantasy to expect the masses to transition from Scratch (or Snap!) to what we now think of as conventional programming languages such as C++, Python or Java. A more plausible future is for specialized languages to spring up that make use of the conventions of Scratch/Snap! but incorporate the knowledge and concepts of a field of interest to a target audience. It isn't difficult to see a world where almost everyone speaks Scratch (like almost everyone knows algebra), and when we go into, for example, molecular biology, we naturally adopt Scratch/Genome, or if one wants to go into accounting, she adopts Scratch/Accounting. In other words, a discipline specific toolbox or plugin will tailor how the programming environment would interface with its users. Keeping these in mind, in this paper, we embark upon describing the contours of a customizable high level programming environment called {\em Patch} using a computing example. Patch is planned to be embedded within a fully autonomous tutoring and assessment system called Mind{\em Reader} \cite{MindReader-ICALT-2017s} for online self-learning.

\section{Patch Highlights using an Example}

High school and freshman CS students learn several fundamental algorithms that they use as building blocks for more complex algorithms they will build in future by combining the basic ones in some fashion. These algorithms are language independent and can be understood at various levels of abstractions. A bubble sorting algorithm, for example, arranges the values of objects in a collection (a list) in order of their significance -- size of their values, shades of their color, or scale of the sounds. While the idea of sorting is simple, developing an algorithm or a computer program for them often is not. Yet, if given a few elements, students can usually rearrange them in the expected order by hand or visually. Decades of research helped develop numerous sorting algorithms that are vital to computing in general and all computer scientists are expected to master them. Figure \ref{sorting} shows the pseudocode of the bubble sort algorithm and its actual C++ implementation. The way the algorithm works, i.e., its principal mechanism, is depicted visually in Figure \ref{bubble} on a list of six values {\tt 29, -4, 2, 17, 45}, and {\tt 9}, as presented to the system, say, by a student.

\begin{figure}[h]
\centering{\begin{tabular}{cc}
\begin{minipage}{.45\textwidth}
{\scriptsize
\IncMargin{1em}
\begin{algorithm}[H]
\KwIn{A list of $n$ values in random order}
\KwOut{Ascending order list}
set $sorted=false$\;
\While{not sorted}{
set $sorted=true$\;
\ForEach{element $i=1,\ldots, n-1$}{
\If{element $i$ $<$ element $i+1$}{
swap elements $i$ and $i+1$\;
set $sorted=false$\;
}}}
\caption{Bubble sort}
\label{alg:bubble}
\end{algorithm}
\DecMargin{1em}
}
\end{minipage} &
\begin{minipage}{.45\textwidth} %
{\footnotesize \begin{verbatim}
void bubbleSort(int ar[]) {
  for (int i = (ar.length - 1);
                      i >= 0; i--) {
    for (int j = 1; j = i; j++) {
      if (ar[j-1] > ar[j]) {
        int temp = ar[j-1];
        ar[j-1] = ar[j];
        ar[j] = temp;
      }
    }
  }
}
\end{verbatim}}
\end{minipage} \\
(a) Bubble sort pseudocode. & (b) C/C++ implementation.
\end{tabular}}
\caption{Bubble sorting algorithm.}
\label{sorting}
\end{figure}

There are many online sorting algorithm visualization tools available including Toptal \cite{Toptal} and David Galles' data structure visualization library \cite{Galles17}. These viz tools do a pretty good job at offering an intuitive understanding of the underlying working principles of the algorithms and even show the differences how they accomplish identical goals possibly with different performance overheads and under the conditions they do so. Nevertheless, these tools do not necessarily help the students develop the abstraction or the algorithm to perform the sorting. Needless to say that the distance from gleaning the sorting idea from these viz tools to developing the corresponding C++ or Python code is still significant, i.e., it is often difficult for a young student to translate a merge sort visualization to a executable C++ code that will be functionally equivalent.

\subsection{Lowest Level of Abstraction}

At the lowest level of abstraction, students write actual programs in Mind{\em Reader} \cite{MindReader-ICALT-2017s} in languages such as C++, Python or even R. They have tool support, to find correct syntax, locating libraries, linking modular programs, compiling and executing, testing and visualizing their results where appropriate, in similar ways the above viz tools allow. Mind{\em Reader} actually extends automated online tutoring and authentic assessment support to students differently and more powerful manner than the recently proposed Java tutor \cite{MartinPSR17s} that intelligently selects predefined templates to explain errors and how to correct them. However, at this abstraction level, students are actually required to develop codes directly in a language of choice.

\subsection{Abstract Thinking}
Instead, let us assume that we ask the student to write a C++ program to rearrange a list of values $a_i$ satisfying the condition that for all $i$, $1 \leq i \leq (n-1)$, $a_i \leq a_{i+1}$ (or, $a_i \ge a_{i+1}$) holds. We also offer her a graphical tool in which she is able to generate a set of $n$ values that she can also visualize in size proportional to the values. We can then contemplate three levels of abstractions - the lowest level is a C++ interface (as discussed above) in which she can directly enter a code fragment, execute it and see the results, and possibly use a visualization tool to see generated objects rearranged using her code.

\begin{figure*}
\centerline{\includegraphics[height=1.25in,width=\textwidth]{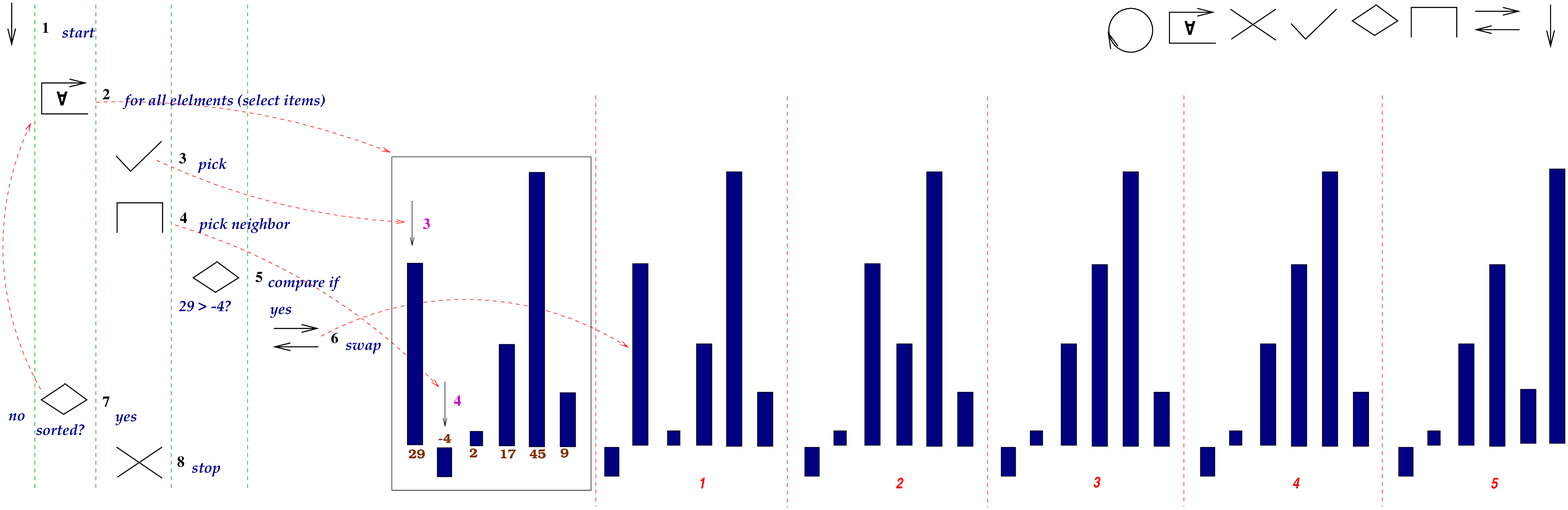}}
\caption{Bubble Sort using Patch (first iteration).} \label{bubble}
\end{figure*}

The highest level of abstraction could be to let her manually and graphically rearrange her objects online. At this level, it is largely conceptual, and the actions will reflect how humans think. For example, she might choose to select the visually largest element at location $j$ to swap position with the one at the $n$th location, or simply bump all the elements up one location up to $j$ to make room at location $n$. If she continues to do so for the next largest, and then the next largest, a pattern will emerge. There are many ways this exercise can be modeled visually and some may be more sophisticated than the others. Regardless, it is possible to settle for one. For example, the visual implementation of the bubble sort algorithms introduced in Figure \ref{sorting} is shown in Figure \ref{bubble}.

In this example, a student used iconized predefined operations on a grid to explain her procedure (numbered 1 through 8) to the computer to help the computer write the code for her in a language of choice. Icons in this figure represent conceptual actions, and further explained using forms, and they are used in a hierarchy -- one to the right on the grid expands the icon immediately on the left, giving it a tree structure. Objects can be selected, highlighted, dragged and dropped to indicate actions. The diagram shows the module begins at step 1, which consists of the remaining steps to the right. She then created a random list of six animated values. In step 2, she selected all the elements to signify that the operations to the right of step 2 apply to all the elements starting with the first one in step 3. She selected the neighbor element in the list, and then compared in step 5, swapped them in step 6 the first being larger than its neighbor. The process can be animated with the push of a button at any point, and be visualized. She noticed that at the end of the animation, the largest element was placed in the sixth position. She then repeated the process in step 2 deciding its not sorted. When she noticed at some point, the list was sorted, she added the stop icon at step 8, completing the algorithm. As we will see in section \ref{patch}, that the level of details needed to implement the algorithm in Figure \ref{sorting}(a) is different, and the description in Figure \ref{patch-int} is more closer to the C++ implementation in Figure \ref{sorting}(b), while the Figure \ref{bubble} is the most abstract.

\subsection{Conceptual Thinking}
To bridge these two levels of abstractions, we envision an intermediate or middle layer, which is still conceptual but uses constructs closer to programming languages such as assignments, loops and decision, and is language and syntax agnostic. Closest analogy to this abstraction layer is a computational biology querying system called VisFlow \cite{VisFlow-BIBM-2016s}. In VisFlow, users express their computational procedures visually using high-level conceptual icons, and connecting them in meaningful ways. While the objectives and functionalities of VisFlow are significantly different from the Patch interface we are proposing, the look and feel, and the approach to expressing algorithms are similar.

\section{Programming in Patch}

The design of Patch is based on the principle of simplicity, clarity and naturalness of computational thinking toward programming with support for modularity and extensibility. These choices roughly means simpler scope rules, structured concept generation and incremental programming, and automatic concept matching for module identification between modules. At the level of language, it also means Patch supports both name independence (similar to C++, Prolog), and position independence (similar to SQL) so that the order or the naming of parameters of a module will have almost no relevance to its appropriateness when called from or used in another module. All a user will have to know what a module conceptually requires as inputs to generate a desired output. In this sense, Patch is more declarative than procedural in design but procedural in its execution.

\subsection{Basic Constructs}

Patch supports an array of simple data types, programming construct, input-output functions and user defined operations keeping simplicity in mind.

\subsubsection{Data Types and Variables}
\label{dt}

Patch supports basic data types integer (e.g., {\tt 14, 2, 657}), real (e.g., {\tt 2.09, 77.21}), Boolean ({\tt TRUE} and {\tt FALSE}) and string (e.g., {\tt "language", "Programming in Patch"}), and native complex types lists (e.g., {\tt [20, 9, 34], [2.3, 50.77], ["Moscow", "Java"]}), sets ({\tt \{87.2, 2.87\}, \linebreak \{"Patch", "Java", "C"\}}) and tuples (e.g., {\tt $\langle$2, "Main Road", "New York", 10026$\rangle$}) of basic types. Except tuple types, all complex types are homogeneous (of the same type). Complex types, however, can have elements of complex types giving endless possibilities, e.g., lists of lists or tuples. Naming a value of basic or complex type is simple -- variable names are case insensitive. Members or elements in a ordered complex type can be accessed using its position in the order using referencing operation {\tt []}. This rule applies to both lists and tuples. Given,
\begin{quote}
{\tt x=["Moscow", "Java", "Pea"]}; and\\
{\tt y=$\langle$2, "Main Road", "New York", 10026$\rangle$,}
\end{quote}
Patch returns {\tt "Pea"} and {\tt 10026} respectively for {\tt x[3]} and {\tt y[4]}, and yes, the base of all collection types start at {\tt 1}. While individuals in lists and set types are not named (being of the same type), members of a tuple type are always named (being potentially of different types and thus conceptually different).

As we just mentioned, they can be accessed or referred by their position, they can also be accessed or referred to by their name. So, both {\tt y[4]} and {\tt y.zip} refer to the member {\tt 10026}, if indeed it is named {\tt zip}. Needless to mention that membership in a set type data can only be tested, and not accessed. Set operations allowed are membership ($\in$), difference ($\setminus$), union ($\cup$), intersection ($\cap$) and cartesian product ($\times$) with their usual mathematical meanings. In the same vein, Patch supports arithmetic operations plus ($+$), minus ($-$), multiplication ($*$), division ($/$), and exponent (\textasciicircum), and comparators less than ($<$), greater than ($>$), equal to ($=$), less than or equal to ($\leq$), greater than or equal to ($\geq$), and logical connectives {\tt AND} ($\wedge$), {\tt OR} ($\vee$), and {\tt NOT} ($\neg$). These operations and comparators are defined over compatible types. Both real and integers are compatible, strings are compatible with strings, and Booleans are compatible only with Booleans. Finally, since Patch is a visual language, a corresponding keyboard syntax for all the above are not truly necessary as we will see.

\subsubsection{Algorithmic Constructs}

In Patch, algorithms are considered logical operations constructed in steps, and by reusing previously defined computationally compatible modules or units. Patch supports five basic operations -- assignment or transformation, branching, and repetition, plus input output functions such as reading and displaying. Technically, these steps are modeled as embedded trees -- steps may include other steps. For example, a repeat operation may include two operations, an assignment and another repetition whereon the second repetition contains two more operations and so on. These constructs are described next.

\paragraph{Assignments and Transformations}

In Patch, computing an expression as a value is treated differently than assigning a value to a variable. Thus assigning constant {\tt 17}, or a variable {\tt y} which as a value {\tt 17}, to a variable {\tt x} is considered assignment. Computing {\tt 16+1} or {\tt y-1} and assigning the value to {\tt x} is considered a computation or transformation because transformations mutate objects, assignments do not. This distinction plays a role in understanding code segments in Mind{\em Reader} tutoring and assessment system of which Patch is a component. In any case, every assignment or transformation is assigned to a variable of appropriate type over which the operations are valid. Thus, {\tt 2+3.57} producing {\tt 5.57} or {\tt 45+3} producing {\tt 48} assigns {\tt 5} to integer {\tt x} stores {\tt 5} and {\tt 48.0} into real {\tt x}, respectively.

\paragraph{Branching}

Branching controls flow of execution which is usually downwards, and is based on some form of Boolean decision. There are three forms of branching -- {\em by-pass, either-or}, and {\em labeled}.
\begin{description}
\item In a by-pass branching (traditional {\em if-then}), a Boolean decision is made if a set of steps should be followed or by-passed to the steps following them.
\item In contrast to a by-pass branching, exactly one set of steps is executed based on a positive or negative decision (traditional {\em if-then-else}).
\item But in a labeled branching, a set of steps is executed from a list of labeled steps (C++ {\em switch}). But unlike languages such as C++ or Java, in Patch, these labels can be any constant matching a variable, i.e., need not be an ordinal.
\end{description}
The branching decisions are based on appropriate Boolean comparisons over Boolean expressions that follow traditional logical expression constructs.

\paragraph{Repetitions or Loops}

There are three basic types of repetitions in Patch -- {\em counter} or {\em automatic, conditional} and {\em sentinel} repetition.
\begin{description}
\item A counter loop or automatic loop (traditional {\em for}) executes its body a fixed number of times, and has a counter or index variable. The counter variable can be assigned a start and an end value before the loop starts, and can be used in side the loop body but cannot be assigned a new value inside the loop body.
\item A conditional loop (traditional {\em while}) continues to execute so long a Boolean condition, called the loop condition, remains TRUE. The loop condition is usually reversed inside the loop body.
\item Sentinel loops continue until an element in a complex data collection, a sentinel or a marker, is found. The collection is usually a list or a set, and its members are used in the loop body one by one until the marker is reached.
\end{description}
In all these loops, as well as in branching, Patch exists the body of steps it is executing if forced using a EXIT command. On exit, control returns to the step immediately after the branch or loop operation and help stop processing under exceptional circumstances.

\paragraph{Modules}

Modules or units have different meanings in various languages and have rich history \cite{Gutknecht89,FlattF98s,AbreuN06s}. We break from traditional approach to modules and procedures to allow flexibility and support intuitive programming and adopt the position that the order of the parameters or their naming should not matter. We use a schema mapping function to establish correspondence instead and allow a call only when a one-to-one mapping between formal parameters can be established. This way, we support both name independence supported in most programming languages and position independence of SQL. This choice though introduces slight uncertainty in module selection, but enhances conceptual understanding. This also means, the traditional function overloading may not work well in Patch. However, in a first computational thinking course/class, we do not see much value in supporting such complex concepts to start with.

\subsection{Reading and Visualization}

Patch supports basic input output operations just as any other language. But additionally to basic display of outputs, Patch also supports powerful animation and live color visualization in ways similar to Protovis \cite{HeerB10} with the objective of supporting comprehension with execution displays similar to Toptal and Galles.

\section{Patch Programming Interface}
\label{patch}

\begin{figure}
\centering
\epsfig{figure=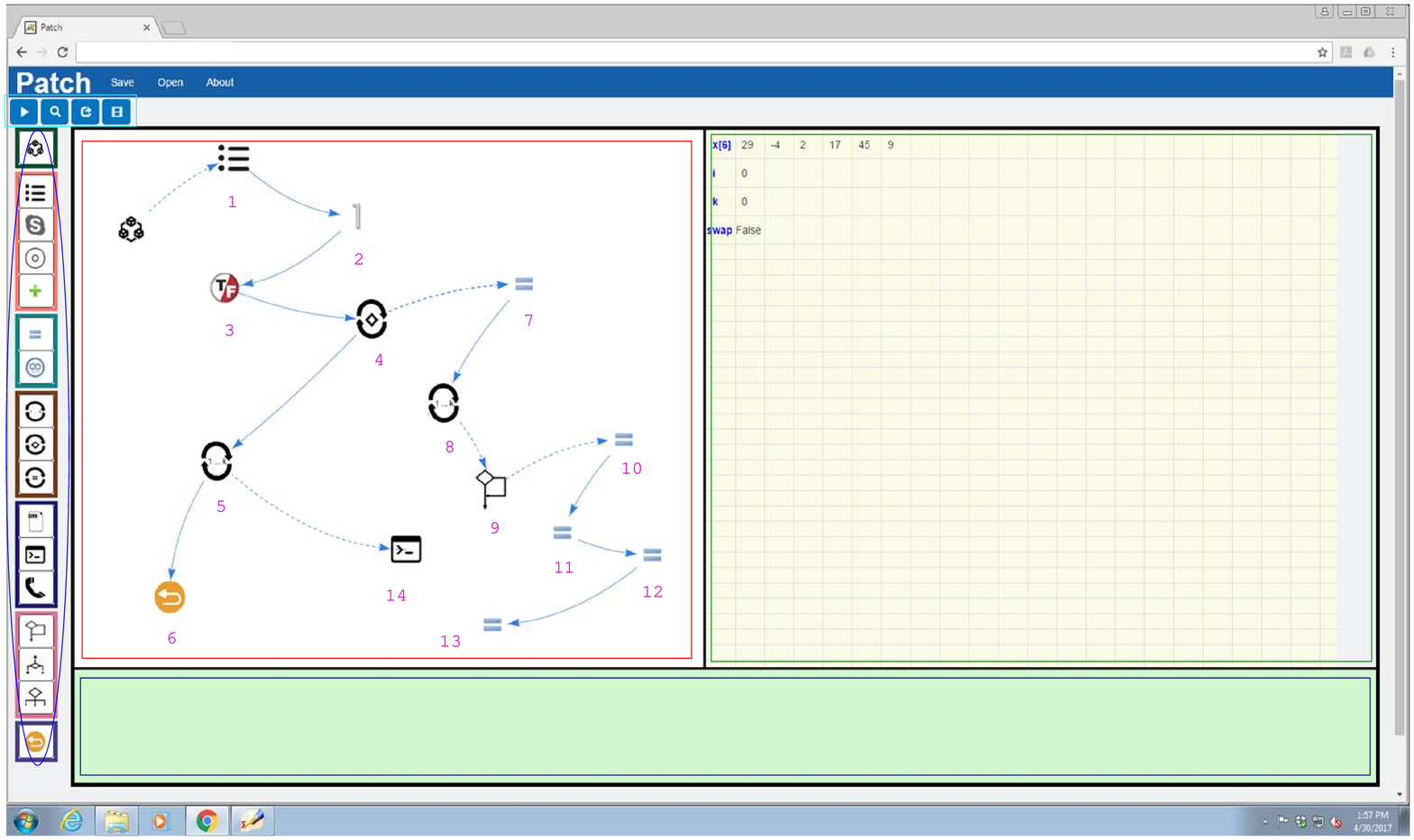,height=2.1in,width=.8\textwidth}
\caption{Patch visual interface showing bubble sort of Figures \ref{sorting}(a) and \ref{bubble}.}
\label{patch-int}
\end{figure}

The Patch programming interface shown in Figure \ref{patch-int} is embedded within the Mind{\em Reader} \cite{MindReader-ICALT-2017s} online tutorial and assessment system. Broadly, the interface has three panels, one icon bank to the left (blue ellipse) and one command and folder bank at the top (cyan box). The left panel (red box) is a canvas where users express their computational solutions, and right panel (green box) serves as the display and input console while the bottom panel (blue box) is reserved for diagnostics and system messages. Users draw a tree involving the icons in the icon bank, the root of which is the module icon. The solid arrows represent downward flow of control or sequence of actions, and the dashed arrows represent membership of actions in another action or step. For example, the dashed arrow from the top module icon means the remaining subtree is contained within the module, and the same way the subtree rooted in 7 is contained within node 4. The Patch tree shown in Figure \ref{patch-int}, is the bubble sort algorithm as designed by a user and Patch is showing the state after step 3.

Technically, every Patch program is a set of named modules each with a (possibly empty) set of input data objects, and a set of data objects it returns when called. Any subset of the data objects in its I/O sets can be from/to the console, a stored repository or another module that uses it. Naturally, I/O from/to console and stored repository takes place inside the module. A module is a pair $\langle D, S\rangle$ where $D$ is a set of data objects it uses, and $S$ is a sequence of logical steps it follows to manipulate the data objects. The four basic, and three complex data types introduced in section \ref{dt} that Patch supports can be used to declare the I/O data objects. Users are able to execute the tree on the canvas at any point and see the results on the green panel. The interface is interactive and actually shows what could be expected on the green panel as the drawing continues. The qualitative details of each node is entered using appropriate forms that can be opened with right clicks of the mouse. The drawing rules ensure all graphs are trees and each node has no more than one solid or one dashed arrow.

\section{Related Research}

The emergence of Scratch and Snap!, and online tools such as Khan Academy or PythonTutor (http://pythontutor.com/cpp.html\#mode=edit) are relatively new. The focus of each of these systems are different and thus often are hard to compare on identical grounds. But we believe in functionality, Patch shares its execution style with PythonTutor, and it is conceptually similar to Scratch and Snap! But Patch combines the power of Mind{\em Reader}'s tutoring and assessment that no other system currently has. We believe the ability to  build programs visually, incrementally and interactively with active tutoring support is a powerful combination. Research is, however, outstanding on the effectiveness of the approach adopted in Patch and Mind{\em Reader} in real life classrooms.

\section{Summary}

Patch is an experimental prototype with limited but emerging set of functionalities and support for multiple abstraction layers for diverse student groups and multiple STEM disciplines. Patch supports exploration and self-learning naturally. It is currently being implemented and tested to validate the hypothesis that all three abstraction layers can be seamlessly supported in a single platform and augmented with an active tutoring and assessment component. Initial experimental results are promising though more research is needed. Patch guarantees generating functionally equivalent code segments at all three of these layers if the solutions specified are indeed identical. Generated and written codes can be compared at the click of a button. Currently, a set of basic programming features for entry level programming course is being identified and abstracted, and methods are being devised to make the system incremental and hierarchical so that complex features can be designed and added using the basic features. A community based approach to incremental design is being contemplated.


%
%
\bibliographystyle{abbrv}
\bibliography{/users/jamil/dropbox/bib-db/bib-db-general,/users/jamil/dropbox/bib-db/our-publications}

\end{document}